\documentstyle[multicol,aps,prl,epsf,psfig]{revtex}
\begin{document}
\title{Steady State Behavior of Mechanically Perturbed
Spin Glasses and Ferromagnets}
\author{David S. Dean and Alexandre Lef\`evre\\
{IRSAMC, Laboratoire de Physique Quantique, Universit\'e Paul Sabatier, 118 route de Narbonne, 31062 Toulouse Cedex 04, France}
}
\date{June 11 2001}
\maketitle

\begin{abstract}
A  zero temperature dynamics of Ising spin glasses and ferromagnets 
on random graphs of finite connectivity is considered, 
like granular media these 
systems have an extensive entropy of metastable states. We consider
the problem of what energy a randomly prepared spin system falls to
before becoming stuck in a metastable state. We then introduce a tapping
mechanism, analogous to that of real experiments on granular media, this 
tapping, corresponding to flipping simultaneously any spin with 
probability $p$, leads to stationary regime with 
a steady state energy $E(p)$. We explicitly
solve this problem for the  one dimensional ferromagnet
and $\pm J$ spin glass and carry out
extensive numerical simulations for spin systems of higher connectivity.
The link with the density of metastable states at fixed energy and
the idea of Edwards that one may construct a thermodynamics with 
a flat measure over metastable states is discussed. 
In addition our simulations
on the ferromagnetic systems reveal a novel first order transition, whereas
the usual thermodynamic transition on these graphs is second order.
\vskip 0.5cm


\end{abstract}
\begin{multicols}{2}
\section{Introduction} 
Recently there has been much experimental and theoretical interest in
the properties of granular media. In such systems the thermal energy
available is not sufficient to allow the rearrangement of a single 
particle and hence the system is effectively at zero temperature in
the thermal sense. The fact that the problem is not trivial lies
in the fact that such systems have an exponentially large number of
such metastable states, which may be also called blocked or jammed 
configurations. Edwards associated an entropy to these configurations
\begin{equation}
 S_{Edw} = \ln(N_{MS}) 
\end{equation}
where $N_{MS}$ is the total number of metastable states of the system
\cite{edw}.
It is reasonable to assume that in complex systems such as granular media
$S_{Edw}$ is extensive meaning that $N_{MS} = \exp(Ns)$ where $s$ 
is the entropy per particle, alternatively one may work with an
entropy per unit of volume which is clearly a more natural choice 
in granular media. Because the system has an extensive number of blocked 
configurations, if it is prepared from a random initial state it will lower
its energy via only energy lowering rearrangements until it becomes
stuck in a metastable state. Normally this first encountered blocked state
will not be that of lowest energy (or most dense packing). In order to change
the state of the system an external perturbation such as tapping or shearing is
required. In between perturbations the system relaxes into new 
configurations. A natural and practically very important question 
concerning this sort of dynamics is: what are the properties of the 
steady state regime obtained via such mechanical perturbation schemes ? 

Recently it has been shown that spin glasses and
ferromagnets on random graphs have  an extensive entropy of metastable
states and  one may  calculate this entropy at fixed
values of the energy \cite{dean2,lefde}. Therefore, though they are 
quite different physically 
to granular materials, these systems have an extensive entropy of metastable 
states as do granular media. The motivation of this paper is to see if one
can understand certain steady state properties of mechanically 
perturbed systems in terms of their organization of metastable states. 
The possibility of using spin glasses as a paradigm for granular material
was first introduced in \cite{edme}.
 
Let us recall an example of an experiment on a system of hard spheres
reported in \cite{exps}. A system of dry hard soda glass spheres
is placed in a glass tube. The system is tapped by using a piston to move
the tube vertically 
through a sine cycle. The tapping parameter $\Gamma$ is defined to 
be the ratio of the maximal acceleration due to the piston 
in the cycle to $g$ the acceleration
due to gravity. After an initial irreversible curve, obtained by increasing
the tapping rate slowly, the system arrives on a reversible curve where
the density is a monotonic function of $\Gamma$, the highest packing
densities being obtained at lowest tapping rate. 
Numerical
simulations on granular media \cite{mehta1} reveal similar behavior
(though the irreversible part of the experimental curve corresponding
to a loosely packed {\em fluffy} metastable state was not
seen).
It was also observed
that at small tapping the relaxation to the final density is extremely
slow and is well fitted by an inverse logarithmic decay of the form
\begin{equation}
\rho(t) = \rho_{\infty} -{\Delta \rho_{\infty} \over1 +B \ln(1 + t/\tau)}
\label{eqlog}
\end{equation}
where $\rho_{\infty}$ (the final density), $\Delta \rho_{\infty}$ , $\tau$
(the characteristic relaxation time) and $B$ are fitting parameters.
It should however be remarked that the behavior of granular systems is
strongly dependent on the tapping mechanism and that horizontal shearing 
\cite{horiz} leads to behavior qualitatively different to vertical tapping.

In this paper we extend and elaborate a preliminary report of 
the results of \cite{delelet}.
The philosophy of the paper is to examine spin glasses
as paradigms for granular media. Here the quantity corresponding
to the density is the energy of the system. We allow the system to
evolve under a random sequential zero temperature single spin flip 
dynamics where
only moves which reduce the energy are allowed. When the system is
blocked we tap it with strength  $p\in [0,1/2]$, that is to say
each spin is flipped with a probability $p$,
the updating at this point being parallel. The system is
then evolved by the zero temperature dynamics until it becomes 
once again stuck, the tapping is then repeated. Physically this
corresponds to assuming that in granular media the relaxation time to
a new metastable state is much shorter than the time between taps.
A similar, though not identical,  tapping
dynamics has also been introduced independently 
in the context of three spin ferromagnetic
interactions on thin hypergraphs  \cite{mehta2}, also in the 
goal of studying the dynamics of granular media. We find that
a stationary regime  is reached
after a sufficiently large number of taps, characterized
by a steady state energy $E(p)$ (analogous to the stationary density
-- the same analogy as used in \cite{mehta2}).
The initial  dynamics from the random initial configuration
into the first metastable state is examined  analytically 
for the one dimensional $\pm J$ spin glass or ferromagnet 
(the two are equivalent by a gauge transformation). We call this 
the initial fall and the average energy of the first metastable
state visited $E_f$ is computed. We then develop a mean field theory
for the dynamics under falling then tapping, interestingly this
theory appears to be exact in the case of the  one
dimensional system and one may calculate $E(p)$ within this scheme, the
results being in excellent agreement with the numerical simulations.

Numerically we examine the tapping of spin glasses and ferromagnets
of higher connectivity. For the spin glass we find that $E(p)$ is, 
as in the experiments, a decreasing function of $p$. For small $p$  we define
the exponent $\theta$ by  $E(p) \sim E(0^+) + A p^\theta$, with $A$ constant. 
In the one dimensional case we show analytically
that $E(p) \sim -1 + \sqrt{2p}$, hence $\theta = 1/2$, whereas for spin
glasses on thin graphs for connectivity superior to two we find that
$\theta = 1$. However for $p <0.05$
we find that the time to reach the steady state is extremely long and 
not accessible numerically. In this slow dynamical regime we find a slow 
relaxation of the time dependent energy, reminiscent of that observed in 
experiments on granular media  \cite{exps} and hence compatible with 
Eq. (\ref{eqlog}).

In the case of the ferromagnet we find numerically that there exists
a critical value $p_c$ of $p$ such that for $p > p_c$,
$E(p) > E_{GS}$ where $E_{GS}$ is the energy of the ground state and
the inequality is strict, and that for $p < p_c$  $E(p) = E_{GS}$. Hence
in the ferromagnetic system there is a first order phase transition
under  tapping dynamics (in contrast to the usual thermodynamic
ferromagnetic transition in these systems which is second order \cite{jon}).
 
There have of course been many models
studied to understand the compaction process in granular 
media \cite{Ben,GM,madegr},
which reproduce many of the experimental features. Here  the spin glass 
is clearly far from a realistic realization of a granular media, however
the fact that it has  extensive entropy of blocked states and the 
obviously natural form of the tapping dynamics implemented makes it
a natural testing ground for ideas about dynamics and possible thermodynamics
of systems such as granular media. Moreover, it has been argued in
\cite{Ben} that the slow compaction regime is well explained if we assume
that particles can rearrange themselves in such a way as to create a 
particle size
void, which is quickly filled by a new grain. This mechanism involves
crossing of energy barriers and leads to a logarithmic compaction before
the  asymptotic steady state regime \cite{GM}. 
We expect that the local rearrangements which
occur during the tapping dynamics on spin glasses random graphs will 
be lead to behavior analogous 
to the  slow glassy dynamics of systems as granular media.

Of course one would ultimately like to obtain a theoretical understanding 
of the asymptotic, steady state regime of lightly tapped granular media.
Edwards has proposed \cite{edw} 
that a light tapping dynamics on granular type systems leads
to a steady state whose properties are determined by a flat measure over
the blocked or metastable states satisfying the macroscopic constraints 
involved ({\em e.g.} fixed internal energy and compactivity). This idea has
recently attracted much interest and has been examined in the context of 
various models \cite{jorge,nico,cug,ledeod}. In this paper we shall
concentrate simply on the asymptotic energy of the final tapped state,
the study of the dynamics leading to this final regime is deferred for further
investigation \cite{lededyn}. We shall see that the calculation of the 
Edwards entropy as a function of energy gives us a possible explanation
of the first order ferromagnetic transition.

\section{Spin Systems on Thin Graphs}

The models we shall consider are spin systems on random thin graphs.
A random thin graph is a collection of $N$ points, each point being linked to
exactly $c$ of its neighbors, $c$ therefore 
being the connectivity of the graph.
The distribution of metastable states in these systems has been recently
considered in \cite{dean2,lefde}.  
The spin glass/ferromagnet model we shall consider has the Hamiltonian 
\begin{equation}
H = -{1\over 2} \sum_{j\neq i} J_{ij} n_{ij} S_i S_j
\end{equation} 
 where the $S_i$ are Ising spins, $n_{ij}$ is equal to one if the 
sites $i$ and $j$ are connected and zero otherwise. 
The fact that the local connectivity is fixed as $c$ imposes the
local constraints $\sum_{j} n_{ij} = c$, for all sites $i$. 
In the spin glass case
the $J_{ij}$ are taken from a binary distribution where
$J_{ij}= -1$ with probability half and $  J_{ij}= 1$ with probability half.
In the ferromagnetic case $J_{ij} = 1$.
Here we define a metastable state as a spin configuration
where any single spin flip does not increase the
energy of the system. Mathematically the total number of 
these metastable states  is expressed as: 
\begin{equation}
 N_{MS}  = {\rm Tr} \prod_{i=1}^N \theta\left( \sum_{j\neq i} J_{ij}n_{ij} S_i S_j
\right)
\end{equation}
It should be pointed out here that the definition of metastable states
is of course dependent on the dynamics of the system, in contrast 
with micro-states in classical statistical mechanics. Whether or not,
 in certain cases,
the information about the dynamics encoded in the calculation of the entropy
of metastable states is enough to allow one to predict the properties of the 
steady state regime is an open question.

The fact that, in our definition of metastable states, we include the marginal case (where the energy change is
zero) implies that here $\theta(x)$, the 
Heaviside step function, is taken such that $\theta(0) =1$. In the context
of granular media, where friction plays an important role, this is a natural 
choice as one certainly needs a non zero force in order to make a
grain move.
With this definition, the total
number of metastable states of internal energy $E$ per spin is formally
given by
\begin{equation}
 N_{MS}(E)  = {\rm Tr} \prod_{i=1}^N \theta\left( \sum_{j\neq i} J_{ij}n_{ij} S_i S_j
\right) \delta(H - NE)
\end{equation}
The corresponding Edwards entropy per spin, at fixed energy $E$ per spin,
is then given by
\begin{equation}
s_{Edw}(E) = \ln\left(N_{MS}(E)\right)/N 
\end{equation}
\section{The One Dimensional Ferromagnet/Spin Glass}
 We remark that by a gauge transformation
the one dimensional ferromagnet and $\pm J$
 spin glass are equivalent and place ourselves 
for transparency in the context of the ferromagnet. Let us remark that
the zero 
temperature Glauber dynamics of the one dimensional
ferromagnet can be explicitly solved \cite{glauber}, here diffusion
of domain walls occurs and the dynamics does not get blocked. In the
Glauber case one may close the dynamical equations, however here
such a closure scheme does not seem possible. The zero temperature 
Kawasaki dynamics (conserving the total magnetization) 
of the one dimensional Ising model, where the
system can freeze, has been solved in \cite{priv}.

To solve the dynamics of the one dimensional ferromagnet we consider
the dynamics from the point of view of the bonds. We define 
a fault of length $n$ to be a sequence of $n$ neighboring adjacent domain
walls. The zero temperature dynamics takes place within these faults via
the flipping of one of the $n-1$ spins contained between the $n$ domain
walls. We define $I_i(n)$ to be the indicator function that starting from
bond $i$ there are exactly $n$ consecutive domain walls (there being no
domain wall on bond  $i-1$ and no domain wall on bond $n+i$ but all
the intervening bonds have a domain wall).
In the initial 
configuration  we take the probability that 
a given spin is different to its left neighbor (that is to say the 
probability of a domain being between two spins) to be $a$. Hence if 
$a=0$ we have an initially ferromagnetic configuration, if $a =1$ it
is an antiferromagnetic configuration, the case $a =1/2$ corresponds to
a completely random configuration of maximal entropy. The total energy
of the initial configuration is then given by
\begin{equation}
{\cal E}_0 = -N + 2 \sum_i \sum_n I_i(n)
\end{equation}  
as the energy is given by the ground state energy $-N$ plus two times the 
number of domain walls (excitations). Defining $K_0 (n)$ to be the probability
that one has $n$ consecutive domain walls starting from a given site
(hence $K_0(n) = \langle I(n) \rangle$ where $\langle \cdot \rangle$
indicates the average over the initial conditions), with the initial 
conditions introduced above one finds
\begin{equation} 
K_0(n,N) = (1-a)^2 a^n
\end{equation}
(The length of the defaults has a geometric distribution).
The initial energy per site of a configuration generated in this manner is
therefore (using the translational invariance of the system when $N \to 
\infty $) given by 
\begin{equation}
E_0 = -1 + {2\over N}\sum_{n=1}^N K_0(n,N) n
\end{equation}

For the distribution of initial configurations considered here we find
therefore that
\begin{equation}
E_0 = -1 + 2 a
\end{equation}

We define by $\chi(n)$ the average number of isolated domain walls
left by a fault of $n$ consecutive domain walls after the zero temperature
dynamics described above has finished. The final energy of the system
per site $E_f$ is therefore
\begin{equation}
E_f = -1 + {2\over N}\sum_{n=1}^N K_0(n,N) \chi(n) \label{eqef}
\end{equation}

By the definition of the spin
dynamics, domain walls disappear by pairs of two neighboring domain walls.
It is clear that $\chi(1) = 1$, $\chi(2) = 0$ and we set $\chi(0) = 0$.
Within such a fault the dynamics proceeds by flipping one of the $n-1$
spins between the domain walls. By recurrence, after a random flip we obtain
\begin{equation}
\chi(n) = {1\over n-1}\sum_{k= 1}^{n-1} \chi(k-1) + \chi(n-k-1) \label{eq:rec}
\end{equation}
We solve equation (\ref{eq:rec}) by introducing the generating functional
\begin{equation}
g(z) = \sum_{n=1}^\infty \chi(n) z^n
\end{equation}
The resulting equation for $g(z)$ is
\begin{equation}
z {dg\over dz} = \left({2 z^2 \over (1-z)^2} + {1\over z} \right) g(z)
\end{equation}
Solving this with the appropriate boundary conditions one obtains

\begin{equation}
g(z) = {z\exp(-2 z)\over (1-z)^2}
\end{equation}
and substituting this into Eq. (\ref{eqef}) yields
\begin{equation}
E_f = -1 + 2(1-a)^2\sum_{n=1}^\infty  \chi(n) { a^n} = -1 + 2(1-a)^2
g(a)
\end{equation}
thus giving the result
\begin{equation}\label{ef}
E_f = -1 +  2 a \exp(-2a)
\end{equation}

This yields a value of $E_f$ for the completely random initial configuration
where, $a = 1/2$, of $-0.632121$.  In fact the value of 
$E_f$ is maximal for the case $a = 1/2$. For the totally antiferromagnetic 
initial 
condition, where $a = 1$, here we find $E_f = -0.72933$. 
Clearly when $a =0$ the system is already in its ground state and we 
find $E_f = -1$ as we should. We note that these values (and those
for all $a$) have been checked with and are in perfect agreement with
our numerical simulations. 

This calculation with the one dimensional ferromagnet demonstrates two
important points:
\begin{itemize}
\item The final value of the energy $E_f$ depends strongly on the initial 
configuration.

\item The system does not fall into a state of energy corresponding
to the maximum of $N_{MS}(E)$. In \cite{dean2,lefde} it was shown
that $N_{MS}(E) \sim \exp(Ns(E))$ where $s(E)$ is a concave function
peaked at $E^* = -1/\sqrt{5} \approx 0.44721$. Hence even if the total number
of metastable states is dominated (in the statistical sense)
by those of energy $E^*$, generic initial
conditions always seem to lead to an energy lower than this
\cite{mehta2,jorge}. In \cite{par} the value of $E_f$ for a variety
of zero temperature dynamics (sequential, greedy and reluctant)
in the fully connected Sherrington Kirkpatrick (SK) spin glass 
model \cite{sk} was studied, similar behavior was found.
 
\end{itemize}

\begin{figure}
\narrowtext
\epsfxsize=0.8\hsize
\epsfbox{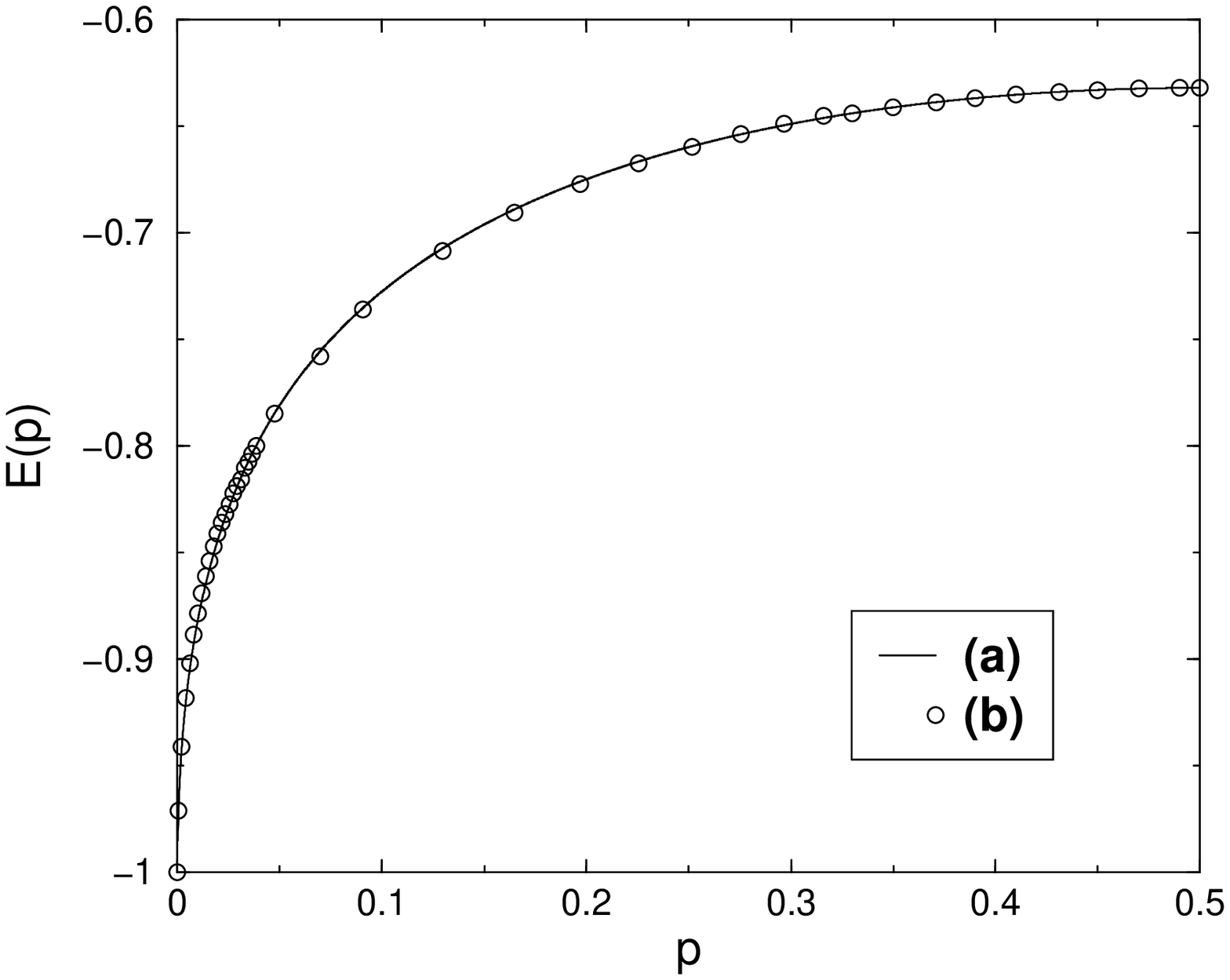}
\caption{}
\label{fig1}
\end{figure}

When the one dimensional system is tapped we find results in
line with those described later for the spin glasses at higher connectivity. 
The curve of $E(p)$, the asymptotic stationary value of the 
energy at a given $p$, is shown in Fig.(\ref{fig1}) from $100$
systems of size $10000$ spins. 

The time taken to reach a stationary value for $E(p)$ were rapid
for larger $p$ but for small $p$ there is a very slow relaxation
to the final asymptotic state which is of the form $1/\sqrt{t}$, where
$t$ is the number of taps. This is easily understood as at very slight tapping
order $p$ effects dominate at early times, this means that:
(i) Isolated pairs of domain walls within large domains
are immediately destroyed once tapping is stopped. (ii) Flipping a spin
either side of a domain wall creates domain wall diffusion and with this
annihilation by coalescence of domain walls. Hence the dynamics at
small $p$ and early times is qualitatively the same as that for the 
low temperature Ising model coarsening \cite{bray}.  

In order to go beyond our first calculation of $E_f$ and solve the 
tapping dynamics we consider a mean field theory for the dynamics
of a system of connectivity $c$. We shall see that at $c=2$ this 
theory gives the analytic result (\ref{ef}) and reproduces 
to within numerical errors the numerical tapping results.
Once again we concentrate on the dynamics on the bonds. For a given site
define $x$ to be the difference between the number of unsatisfied and 
satisfied bonds. Hence $x$ is the local field on the spin at this site and
$x\in -c, \ -c+2, \cdots, c-2,\ c$. If $x > 0$ then the spin can flip 
bringing about the change $x\to -x$. 
In addition we denote by $P(x,k)$ the probability that
the site of interest has local field $x$ after a total of $k$ 
attempted random sequential spin flips under the zero temperature falling 
dynamics (that is to say the dynamics in between taps).
We define $f_+$ and $f_-$ the probabilities that a given spin 
can flip conditional on the fact that the bond with a given neighboring site 
is not satisfied or satisfied respectively. Formally we have 

\begin{equation}
f_\pm = {\rm Prob}\left(x >0 \vert \hbox{given bond is not satisfied/satisfied}\right)  \label{eqfpm}
\end{equation}

We may turn around this conditional probability using Bayes' Theorem to obtain

\begin{equation}
f_\pm = {{\rm Prob}\left(x >0\  \hbox{and  given bond is not 
satisfied/satisfied}\right)  
\over {\rm Prob}\left( \hbox{ given bond is not satisfied/satisfied}\right)}
\end{equation}
Given that a site has local field $x$, it must have $(c+x)/2$ unsatisfied bonds
and $(c-x)/2$ satisfied bonds. Therefore we find that
\begin{eqnarray}
& &{\rm Prob}\left(x > 0 \   \hbox{and given bond is not
 satisfied/satisfied}\right) \nonumber \cr
 &=& \sum_{x>0}P(x)(c\pm x)/2c
\end{eqnarray}
and
\begin{eqnarray}
& &{\rm Prob}\left( \hbox{given bond is not satisfied/satisfied}\right)
\nonumber \cr
&=& \sum_x P(x)(c\pm x)/2c
\end{eqnarray}
Putting these results into Eq. (\ref{eqfpm}) then gives
\begin{equation}
f_\pm  = {\sum_{ x> 0} P(x)(c\pm x) \over \sum_x P(x)(c\pm x)}
\end{equation}

If we are interested in the spin at site $i$ the possibilities between time $k$ 
and $k+1$ are
\begin{itemize}
\item The spin at site $i$ is chosen and 
$x>0$, then the spin at site $i$ will flip and $x$ goes to
$-x$.
\item The spin at site $i$ is chosen and 
$x \leq 0$, then the spin at site $i$ can not  flip and $x$ 
does not change.
\item  A neighbor of site $i$ with  positive local field 
is chosen and
so  flips. In this case, $x$ goes to $x+2$ or to $x-2$ depending whether
or not the bond with site $i$ was satisfied or not satisfied.
\item A neighbor of site $i$ with negative or zero  local field 
is chosen and  so does not
flip. In this case, $x$ does not change
\item One  chooses neither the spin at site $i$ nor any of
its neighbors, and so $x$ stays $x$. 
\end{itemize}
Assuming that the distribution at every site is given by $P(x,k)$ and 
assuming independence between the values of $x$ from site to site (the
mean field approximation) we obtain

\begin{eqnarray}\label{eq}
P(x,k+1)&=&\frac{\theta(-x)}{N}P(-x,k)+\frac{\theta(-x)}{N}P(x,k)
 \nonumber \\
&+&\frac{N-c-1}{N}P(x,k) \nonumber \\
        &+&P(x,k)\,\left(\frac{c+x}{2N}(1-f_+)
+\frac{c-x}{2N}(1-f_-)\right)  \nonumber \\
	&+&P(x+2,k)\,\frac{c+x+2}{2N}f_+ \nonumber \\
&+&P(x-2,k)\,\frac{c-x+2}{2N}f_-
\end{eqnarray}
In this equation we have to define $\theta(0)$, the choice compatible with
the conservation of probability is $\theta(0)=1/2$.
Taking the limit $N \to \infty$ we may introduce the continuous time
$\tau = k/N$ and obtain
\begin{eqnarray}\label{eqn}
{dP(x)\over d\tau}&=& \theta(-x)P(-x)+\theta(-x)P(x)
-(c+1)P(x) \\ \nonumber
        &+&P(x)\,\left(\frac{c+x}{2}(1-f_+)
+\frac{c-x}{2}(1-f_-)\right)\\ \nonumber
	&+&P(x+2)\,\frac{c+x+2}{2}f_++P(x-2)\,\frac{c-x+2}{2}f_-
\end{eqnarray}
The average energy per site at time $\tau$ is then given by
$E(\tau) = {1\over 2}\sum_x xP(x,\tau)$ and one can show that the above
mean field equation (\ref{eqn}) respects the exact identity for the evolution
of the average energy per spin 
\begin{equation}
{dE \over d\tau} = -2\sum_{x>0} x P(x,\tau).
\end{equation}
The case where $c=2$ (the one dimensional case) is accessible to analytic 
solution and we proceed by defining

\begin{eqnarray*}
u(\tau) &\equiv& P(-2,\tau) \\
v(\tau) &\equiv& P(0,\tau) \\ 
w(\tau) &\equiv& P(2,\tau) 
\end{eqnarray*}

One finds that

\begin{eqnarray}\label{eqn1}
f_-&=&0\\ \nonumber
f_+&=&\frac{2w}{v+2w}
\end{eqnarray}
 and the full mean field evolution equations become
\begin{eqnarray} \label{eqn2}
\frac{du}{d\tau}&=&f_+ v+w \\ \nonumber 
\frac{dv}{d\tau}&=&-f_+v+2f_+w \\ \nonumber 
\frac{dw}{d\tau}&=&-(2f_++1)w
\end{eqnarray}

If we look for a stationary solution of (\ref{eqn1}) and (\ref{eqn2}),
we find $w=0$, which expresses the fact that when $\tau$ is infinite, 
the system is in a metastable state. To solve (\ref{eqn1}) and
(\ref{eqn2}), we introduce $\lambda$ so that $v=\lambda w$. Then
$\lambda$ obeys the equation:
\begin{equation}
\frac{d\lambda}{d\tau}=\lambda+2,
\end{equation}
with the initial condition $\lambda(0) = v(0)/w(0)$.
Then (\ref{eqn2}) becomes:

\begin{eqnarray} \label{eqn3}
\frac{du}{d\tau}&=&\frac{2\lambda w}{\lambda+2}+w \\ \nonumber  
\frac{dw}{d\tau}&=&-\frac{\lambda+6}{\lambda+2}w \\
\lambda+2 &=&(\lambda(0)+2)\, e^\tau
\end{eqnarray}
and $w(\tau)$ and $v(\tau)$ are given by

\begin{eqnarray}
w(\tau)&=&w(0)\,\exp\left(-\tau+\frac{4}{\lambda(0)+2}(e^{-\tau}-1)\right)\\
v(\tau)&=&-2w(\tau)  \nonumber \\ &+& w(0)(\lambda(0)+2)\,
\exp\left(\frac{4}{\lambda(0)+2}(e^{-\tau}-1)\right)
\end{eqnarray}
The probability to have a positive value for the local fields then goes to
zero at infinite $\tau$ as expected and the limit of $v$ is 
$v(\infty)=(v(0)+2w(0))\,e^{-\frac{4}{\lambda(0)+2}}$.
If we consider the geometric initial conditions used in the previous exact
calculation of $E_f$,
the induced initial conditions are: $u(0)=(1-a)^2$, 
$v(0)=2a(1-a)$ and $w(0)=a^2$.
In this case we obtain 
$E_f=-1+v(\infty)=-1+2a e^{-2a}$ reproducing the exact result (\ref{ef}).
Tapping the system with tapping probability $p$, starting from the 
values 
$\{u(\infty),v(\infty),w(\infty) \}$, we obtain the new {\em tapped} values 
$\{u^\prime(0),v^\prime(0),w^\prime(0) \}$. Defining $q \equiv
(1-p)$, the relations between the old and {\em tapped} probabilities are:

\begin{eqnarray}
u^\prime(0)&=&(1-3pq)\,u(\infty)+pq\,v(\infty) \\ \nonumber
v^\prime(0)&=&2pq\,u(\infty)+(1-2pq)\,v(\infty) \\ \nonumber
w^\prime(0)&=&pq
\end{eqnarray}
Then, after another zero temperature evolution of the system, it reaches
a new local energy probability distribution with $w^\prime(\infty)=0$ and:

\begin{eqnarray}
v^\prime(\infty)&=&(4pq+v(\infty)(1-4pq)) \nonumber \\
&\times& \exp\left(-\frac{4pq}{4pq+v(\infty)(1-4pq)}\right)
\label{eqn4o}
\end{eqnarray}

At this stage of computation one should remark that this recursive equation
contains one of the main features of our numerical simulations, that is
{\em reversibility}. Indeed, the process involved in (\ref{eqn4o}) will
reach an
asymptotic value which is independent of the initial conditions and depends
on $p$. Hence in the steady state regime under tapping, the 
probability $v_{s}(p)$ (the subscript {\em s} indicating steady state)
for  sites to have  zero local field 
is solution of the fixed-point equation:

\begin{eqnarray}
v_{s}(p) &=& (4pq+v_{s}(p)(1-4pq))\nonumber \\
&\times& \exp\left(-\frac{4pq}{4pq+v_{s}(p)(1-4pq)}\right) \label{eqn4}
\end{eqnarray}
This equation can be solved numerically and the result is shown in 
Fig. (\ref{fig1}) in
comparison with the numerical simulations which we see is excellent.
 
The small $p$ behavior of $E(p)$ from (\ref{eqn4}) is:
$E(p)=-1+\sqrt{2p}+O(p)$, indicating that in this case $\theta = 1/2$.

Given the mean field nature of the above calculation we have used we do
not expect this approximation  to correctly  describe the approach towards 
the steady state, by direct comparison with the numerical simulations we have 
verified that this is indeed the case. Let us remark here that  a 
defect of the mean field approximation scheme is that it cannot distinguish
between a spin glass and a ferromagnet, this is clearly not a problem
for the one dimensional situation where the two are identical. 

\section{Higher connectivities}
The systems which we study are $\pm J$ spin glasses or 
uniform ferromagnets on random graphs
with fixed connectivity $c$. Let us first recall some analytical results
of \cite{dean2,lefde}. It has been 
found that the mean number of metastable states
 increases exponentially with the number of sites in both cases. 
In addition in 
\cite{lefde} an annealed approximation to the Edwards entropy per spin of
metastable states at fixed energy $E$ was carried out: 
\begin{equation}
s_{Edw}(E) = \ln\left( \langle N_{MS}(E)\rangle\right)/N
\end{equation}
which may be exact 
for the ferromagnet as 
the calculated entropy is always positive. 
Moreover, there is an energy threshold $E^*$ above which the results
are the same for the $\pm J$ spin glass and the ferromagnet;
below  the ferromagnet has more metastable
states and a non-zero magnetization. Hence, as far as the energy density
of metastable states is
concerned, both ferromagnet and spin glass are the same above $E^*$ -- that is
the effect of loop frustration is negligible. In this regime, one  also 
suspects that
the zero temperature dynamics are the same. In
particular, numerical simulations with $100$ samples of $N=10000$ sites
for connectivities of $3$, $4$ and $5$
have found the same $E_f$ for the spin glass and ferromagnet
with very good accuracy (the relative error is
about $10^{-6}$). The results are show in Table (1).

\begin{figure}
\narrowtext
\epsfxsize=0.8\hsize
\epsfbox{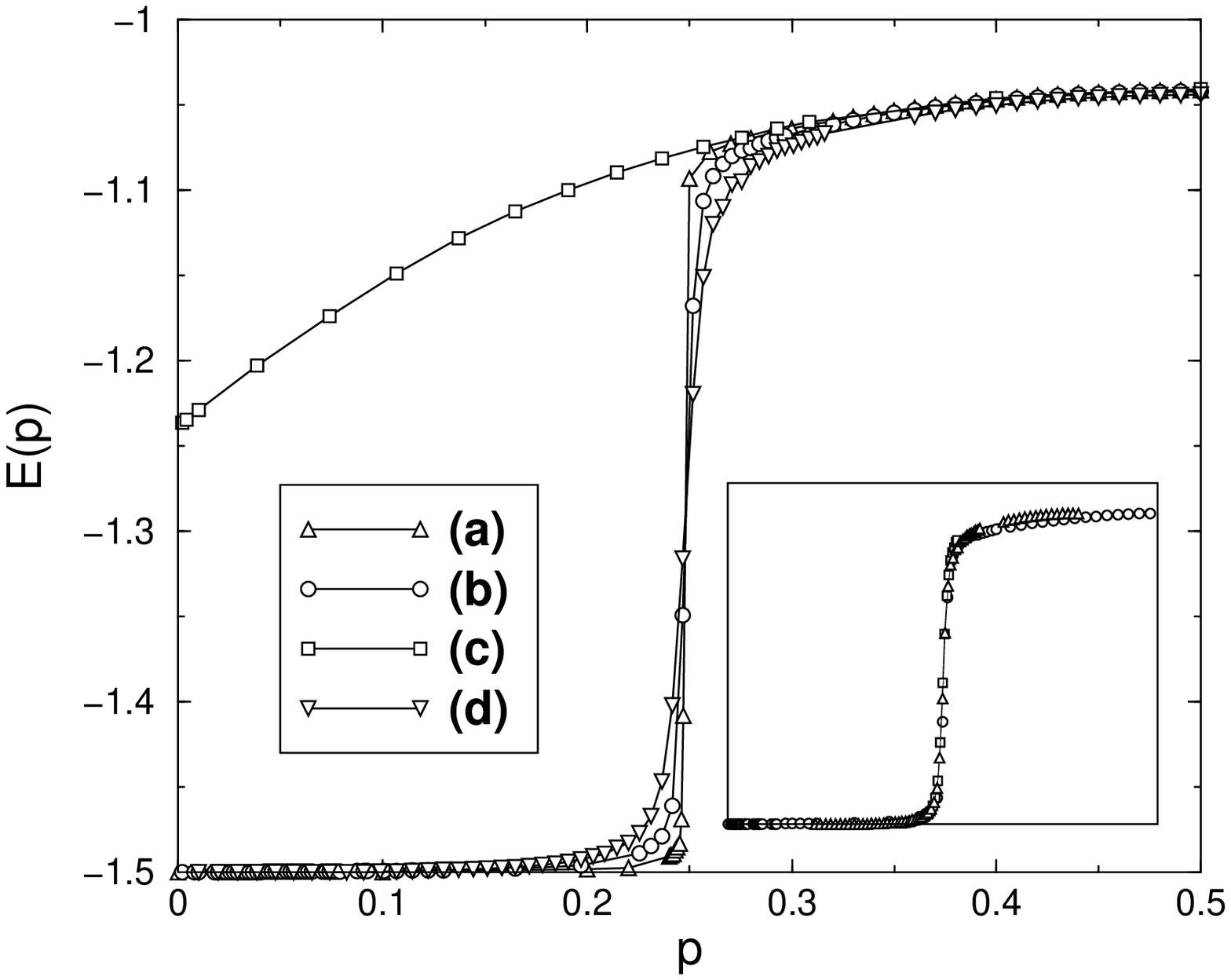}
\caption{}
\label{fig2}
\end{figure}

\begin{figure}
\narrowtext
\epsfxsize=0.8\hsize
\epsfbox{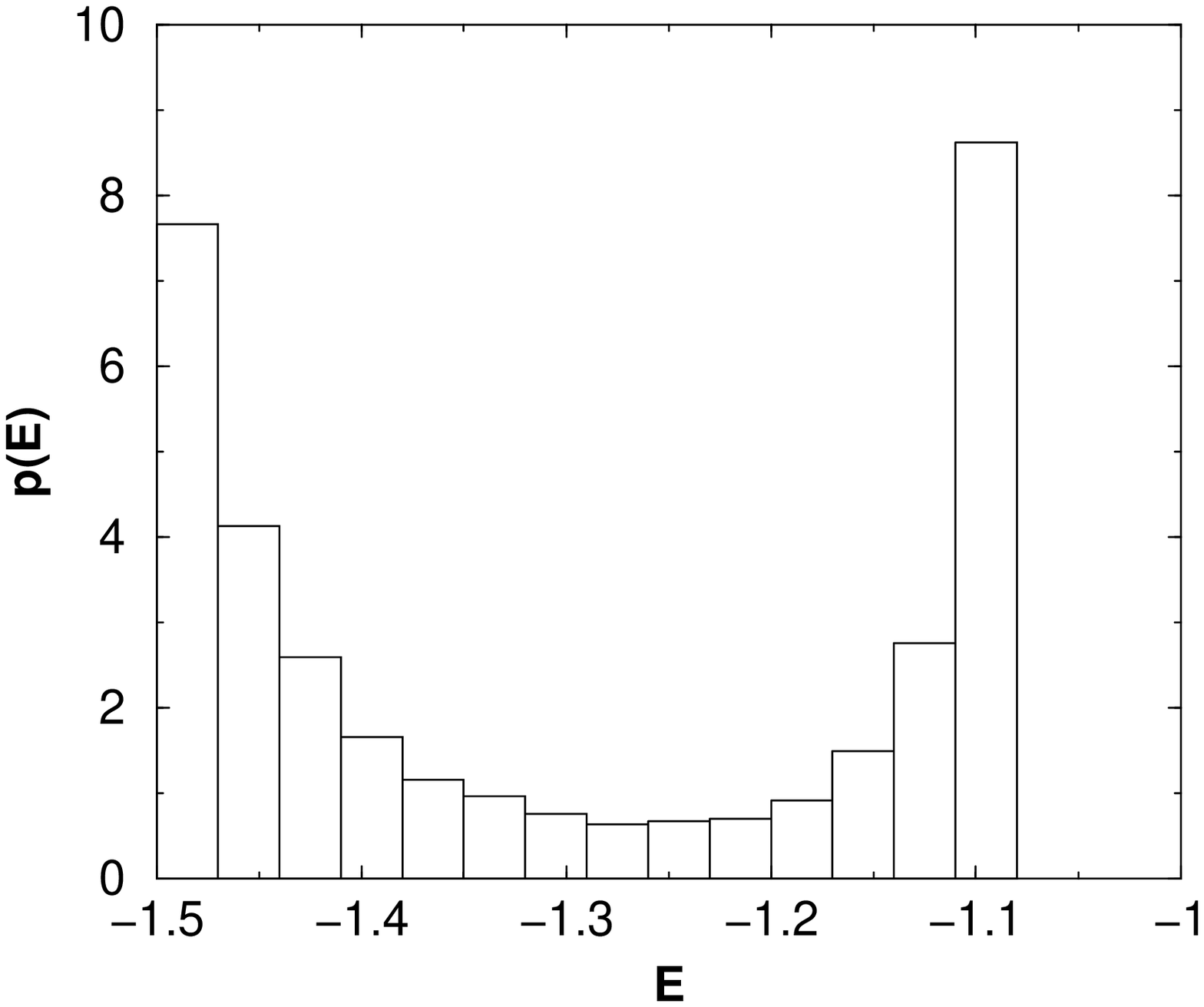}
\caption{}
\label{fig3}
\end{figure}

The result of tapping experiments on the systems with 
$c = 3$ is
displayed in Fig.(\ref{fig2}). There is some critical tapping rate, $p_c$,
above which the curves of $E(p)$ 
versus $p$ are the same for the spin glass and the ferromagnet. 
Moreover, the ferromagnet
is subject to a phase transition under tapping dynamics at $p_c$ such
that for $p<p_c$, the steady state reached is the ground state.  
Finite size effects have been studied and revealed that the transition 
is first order (in as far that $E^{FM}(p_c^+) \neq E^{FM}(p_c^-)$),
in contrast to the usual
thermodynamic ferromagnetic transition in these systems which is second
order \cite{jon}. For the ferromagnet in the region close to $p_c$
one finds an excellent scaling of the energy as a function of $N$,
$E^{FM}(N,p) = f(N(p-p_c))$ as shown in the inset of  Fig. (\ref{fig2}).
This scaling may be used to optimize the determination of $p_c$.
The first order nature of the transition may be seen 
explicitly by looking at the histogram over time (in the steady state regime)
of the average energy per spin at $p = p_c$; 
one sees in Fig. (\ref{fig3}) two separated
peaks in the distribution and not a single peak which splits into two as one 
would expect for a second order transition. Near $p_c$, for systems of 
finite size, there is therefore  coexistence of the two
phases. The time dependence of the average energy per spin in the 
simulation leading to the histogram  Fig. (\ref{fig3})
is shown in  Fig.(\ref{fig4}).
One sees that the system tunnels  between the two coexisting states. The 
typical time for this tunneling increases as the system size increases,
indicating, in thermodynamic language, a {\em free energy} barrier between 
the two phases. As the system size is increased the occupation of the
intermediate states of energy between the two phase of energy $E_{GS}$ and 
$E(p_c^+)$ (between the two peaks in Fig.(\ref{fig3})) is suppressed.

We have also measured  $E(p)$ by studying single systems of very large  size
($N=10^6$), the results are shown in Fig. (\ref{fig5}). 
Here again  above $p_c$ the curve for the spin glass and the
ferromagnet are completely indistinguishable and the 
ferromagnet reaches the ground state
below $p_c$. For such large sizes, 
one no longer sees a coexistence of two phases
around $p_c$ as presumably the tunneling time has become much larger than
the simulation time.

Moreover, for $N=10^6$, for  $p> p_c$ 
the full temporal plots (and not just the steady state values)
of $E^{SG}(p,t)$ and
$E^{FM}(p,t)$ (where $t$ is the number of taps) are indistinguishable.
For $p<p _c$ the two curves are identical  up till a time
$t_{dif}$, which depends on the initial configuration and the sequence of
spins flipped during the tapping process, and diverge after $t_{dif}$, when
the ferromagnetic system reaches quickly the ground state 
(see Fig.(\ref{fig6})). Once the ferromagnetic system has broken 
the $Z_2$ symmetry
the easiest way to lower the energy is to flip the spins 
which are opposed to the global magnetization (because they are more probable
not to be in the direction of their local field) 
until all the spins are $-1$ or $+1$.

Identical behavior was found in the systems with $c=4$ and $5$. 
The comparison of $E^{SG}(p)$ and $E^{FM}(p)$ for  $c=4$ is 
shown in Fig. (\ref{fig5}). Remark that if we compare the different values
of $p_c$ when increasing $c$, and considering only odd (or even)
connectivities, we find that $p_c$ grows, and we expect that it goes to
$1/2$ when $c$ is very large, as the metastable states are more and more
magnetized when $c$ grows (in the case of the fully connected ferromagnetic
Ising model,  only the two ground states are metastable, so
$p_c=1/2$). 

The behavior of the spin glass systems is similar to that for the 
system with $c=1$. The steady state energy  $E^{SG}(p)$ is a 
monotonically decreasing  and continuous 
function of $p$. For small $p$ one finds that here $E^{SG}(p) \sim
E^{SG}(0) + A p$ giving $\theta = 1$ in contrast with $\theta = 1/2$ in the
one dimensional case.

{\em A tentative explanation for the ferromagnetic transition}:
In \cite{lefde} it was also
shown that for the ferromagnet the Edwards entropy as a function of $E$
is concave for $E>E^*$ and convex for $E<E^*$. The value of $E(p_c^+)$
obtained from the tapping experiments are very close to those 
obtained for $E^*$ in \cite{lefde}, the energy at which $s_{Edw}$ becomes 
convex. The results are shown in Table (1). Encouraged by this
striking observation we will try to make a tentative link with a 
possible thermodynamics for such systems. If we imagine that the 
energy of the system is governed by a partition function inspired by the 
flat Edwards measure over metastable states \cite{edw,nico}
\begin{equation}
Z = \int dE N_{MS}(E) \exp(-N \beta E)
\end{equation}
where $\beta$ is a Lagrange multiplier corresponding to the inverse Edwards
temperature which depends solely on $p$ and not on $E$ and is a monotonically
decreasing function of $p$ for $p \in [0^+, 1/2]$. The monotonicity 
hypothesis is supported by the simulation results that $E(p)$ decreases with
decreasing $p$. Clearly the energy which dominates in the sum is that obeying
\begin{equation}
 {\partial s_{Edw}(E)\over \partial E} -\beta = 0
\end{equation}
However if this saddle gives a true maximum of the action one must 
have also that
\begin{equation}
{\partial^2  s_{Edw}(E)\over \partial E^2} < 0
\end{equation}
and hence the Edwards entropy must be concave for the energy
considered to be thermodynamically stable. Hence for $E< E^*$, this suggests 
that the only stable
energy is the ground state.

If one would like to push the analogy of the thermodynamics of first order 
phase transitions  to its limits
and assume that $\beta$ is a continuous function of $p$ (it is clear that
if this is not the case one could trivially obtain a first order transition),
one would also expect that the {\em free energy} of the two phases is
equal at $\beta_c = \beta(p_c)$ and hence 
\begin{equation}
s_{Edw}(E^*) - \beta(p_c) E^* = s_{Edw}(E_{GS}) - \beta(p_c) E_{GS}
\end{equation}
where $E_{GS}$ is the energy of the ground state of the ferromagnet 
given here by $E_{GS} = -c/2$.
It is also clear that $s_{Edw}(E_{GS}) = 0$ hence one obtains
\begin{equation}
\beta(p_c) = {s_{Edw}(E^*) \over E^* - E_{GS}}
\end{equation}
However one also has that
\begin{equation}
\beta(p_c) = \beta(E^*) = {\partial s_{Edw}(E)\over \partial E}\vert_{E = E^*}
\end{equation}
and hence one should obtain
\begin{equation}
{\partial s_{Edw}(E)\over \partial E}\vert_{E = E^*} 
= {s_{Edw}(E^*) \over E^* - E_{GS}} \label{eq:fot}
\end{equation}
However the fact that the calculated annealed approximation
for $s_{Edw}(E)$ is convex for $E<E^*$ means that
\begin{equation} 
{\partial s_{Edw}(E)\over \partial E}\vert_{E = E^*} 
>  {s_{Edw}(E^*) \over E^* - E_{GS}}
\end{equation}
and hence the equality (\ref{eq:fot}) is not respected.
It is possible that the exact quenched calculation  of $s_{Edw}(E)$
would give a different value from the annealed calculation
of \cite{lefde}, however one should not expect the result to be too 
different qualitatively to that of \cite{lefde}. 

In addition we 
remark that the annealed calculation $\ln\left( \langle N_{MS}(E)\rangle
\right)$ gives an upper bound for the
quenched  value $\langle \ln\left( N_{MS}(E)\right)\rangle$  
(from Jensen's inequality). Also the value of $E^*$ found
with this approximation seems to be greater than that found numerically for
$E(p_c^+)$ by a very small amount ($\approx .005$ for $c=3$).
If we accept the possibility that $E(p_c^+)=E^*$, a more
probable hypothesis is the collapse of validity of Edwards' hypothesis in the
region of energy where the $Z_2$ symmetry is broken on metastable states, as
has been found for the three dimensional random field Ising model 
in \cite{jorge}.

Let us mention here that we measured the energy in the simulations 
over a few hundred time steps after the energy appeared to stop to decay.
To be sure that the systems considered here were  in a
stationary regime (and that the energy was
not decaying extremely slowly as a function of time) 
we measured the correlation function at
different waiting times $t_w$ (the number of taps after the initial
preparation of the system), that is to say
\begin{equation}
C(t+t_w,t_w) = {1\over N} \sum_{i=1}^N S_i(t_w) S_i(t_w + t)
\end{equation}
In the stationary regime this should be a function only of $t$. In out of 
equilibrium systems the fact that the system is not
in equilibrium shows up strongly as aging in the correlation function
{\em i.e.} $C(t+t_w,t_w) \neq C(t)$ 
(see \cite{bocukume} and references within), even though
the energy may be decaying so  slowly that it appears to have reached its 
asymptotic equilibrium value. The time translational invariance of 
$C(t+t_w,t_w)$ is thus quite a rigorous test of whether the steady state 
regime has been attained.
For example for the case $p = 0.02$, with the waiting times 
$t_w = 30000$ and $t_w = 60000$
is shown in Fig. (\ref{fig7}), one sees, clearly that after the appropriate
translation of the $t$ axis, the two functions collapse perfectly
 onto one another.
One also sees that the decay of
$C(t)$ (in the longtime regime we can now eliminate the $t_w$ dependence)
is exponential at large $t$ and also that $C(t)$ decays to zero, indicating a
form of ergodicity in the system. As pointed out in \cite{jorge}, this
behavior of the correlation function seems a  necessary condition for the
validity of the scenario of Edwards, that under tapping all metastable states 
satisfying the relevant macroscopic constraints (fixed 
energy and compactivity) are equiprobable in the stationary
regime of gently tapped or perturbed system.

In order to test the accuracy of the mean field approximation
at high energies (where the system does not distinguish between the
ferromagnet and the spin glass), we have compared the  value of 
$E_f$ obtained in the numerical simulations with the result $E_f^{MF}$
obtained by numerical integration of Eq. ( \ref{eqn}). The comparison is in
Table (1) and  we see that the agreement is quite good.

Finally we mention that we  have also examined the reversibility of 
the tapping mechanism. 
If the system
is tapped for a sufficiently long time, compatible with the relaxation times
discussed above, the system is completely reversible. This reversibility was
found in the experiments in \cite{exps} once the system had left the 
initial fluffy state.

\section{Conclusion}
Granular media are a natural example of systems having an extensive entropy
of metastable states. In such systems the role of thermal fluctuations
are negligible and in order to evolve one must apply some external
tapping mechanism. One would ultimately like to be able to formulate
some sort of thermodynamics for such systems. The proposition of 
Edwards \cite{edw}
for a thermodynamics of such systems is an important step in this
direction and has had some success \cite{jorge,ledeod} but it has been
shown not to be generically true \cite{jorge}. A more
general understanding of the asymptotic states of tapped systems has
far reaching implications for computer science as the tapping
mechanism studied here is similar to certain algorithms used in
optimization problems. 

We have presented what appears to be an 
exact calculation of the steady state energy of a tapped one dimensional 
spin glass or ferromagnet. For this problem we have obtained 
the fixed point equations for the distribution of local fields
under tapping. These equations also explain the 
reversibility observed in the numerical simulations. 
In a wide context of models
we confirm the observations of \cite{exps,mehta1,mehta2}, that if one reduces
the {\em strength} of tapping, then the compaction process, corresponding
here to the reduction of the energy of the system, becomes more efficient.
The existence of a first order type phase transition for tapped
ferromagnets on random thin graphs is of great interest, the 
possible explanation using the calculations of \cite{lefde} on the 
Edwards entropy for this system indicates the possibility that one
may eventually construct a more general theory for the thermodynamics
and even phase transitions in tapped systems. One is tempted to speculate that
generically the convexity of the Edwards entropy below a certain
energy threshold (denoted here by $E^*$) leads to a {\em collapse}
to the ground state energy $E_{GS}$, the metastable states in the intervening
energy values being unable to support a stable thermodynamics. 
In terms of granular media this sort of transition would correspond to a 
transition between a random close packed state to a crystalline close packed
state. It would be interesting to find other systems (both theoretical 
and experimental) showing the same collapse phenomena in order to test
this idea.

Finally let us mention
some of the open questions posed by this study we believe to be of
interest for future investigation. Clearly a general goal would be,
in the spirit of Edwards, to develop a thermodynamics to 
describe the stationary regime of tapped systems such as those
studied here. The exact results presented on one dimensional 
systems here provide a completely analytic understanding of the
tapping dynamics which one may be able to rederive from static
considerations. Indeed it has been shown \cite{ledeod} in this simple context that several
steady state observables may be predicted  using
Edwards' measure. The phase transition found in the case of the
ferromagnetic systems studied here is extremely novel, an analytic
understanding of this phenomena would be desirable, perhaps there
exists a percolation type argument which would allow one to evaluate
$p_c$. Also of interest is the decay of a system towards its final 
steady state energy. The slow logarithmic decay described by
Eq. (\ref{eqlog}) has been used successfully to fit the experimental
data of \cite{exps} and the simulation data of \cite{mehta2,GM}. 
Our preliminary study of finite connectivity spin glasses 
and the SK model \cite{lededyn} indicates the presence of a slow
dynamical regime for small values of the tapping parameter $p$ which is also
compatible with a slow logarithmic decay, but the curves can also be well 
fitted by power law decays  (with the same number of fitting parameters). 
There exist however phenomenological arguments \cite{Ben} and 
exact calculations and simulations 
on toy models \cite{Ben,GM,madegr} which support logarithmic decay.

\end{multicols}

{\bf {Figure Captions}}

Tab. 1. Comparison of the numerical values of $E^*$ and $E(p_c^+)$ for
        different values of the local connectivity c. The result for                   $E^*$ when  $c=3$ is a truncation of the analytical value $-15/14$.

Fig. 1. Comparison between numerical simulations of tapping experiments (b)
        and the analytical result (a) obtained with (\ref{eqn4}).

Fig. 2. Numerical simulations of tapping experiments for the spin glass (c)
        and the ferromagnet ((a), (b) and (d)) for $c=3$ for $N=1000$ 
        ((c) and (d)), $N=2000$ (b) and $N=10000$ (a). The inset shows 
        the scaling $E^{FM}(N,p)=f(N(p-p_c))$ for $p \simeq p_c$ for $N=400$, 
        $N=1000$ and $N=2000$.

\begin{figure}
\narrowtext
\epsfxsize=0.8\hsize
\epsfbox{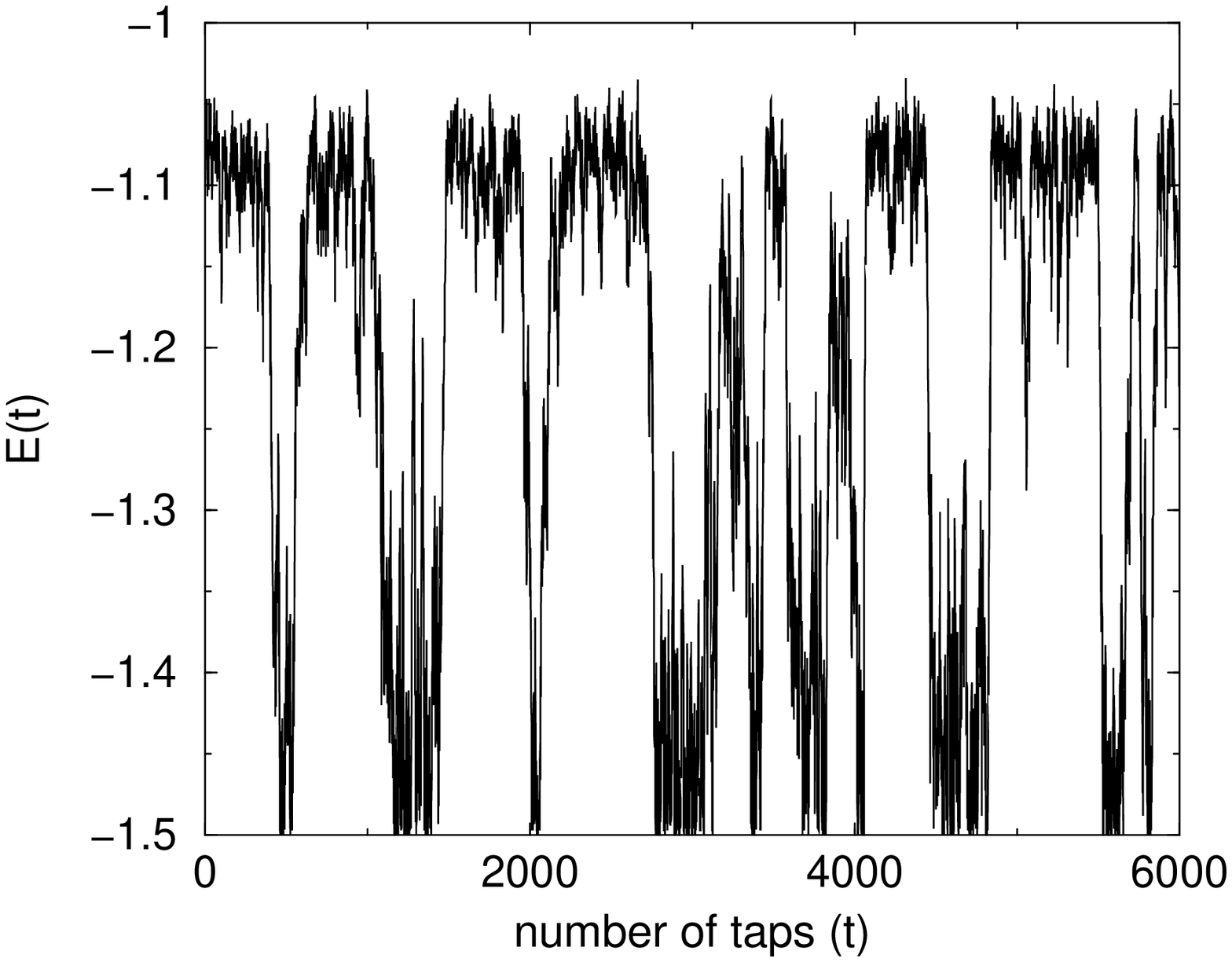}
\caption{}
\label{fig4}
\end{figure}

\begin{figure}
\narrowtext
\epsfxsize=0.8\hsize
\epsfbox{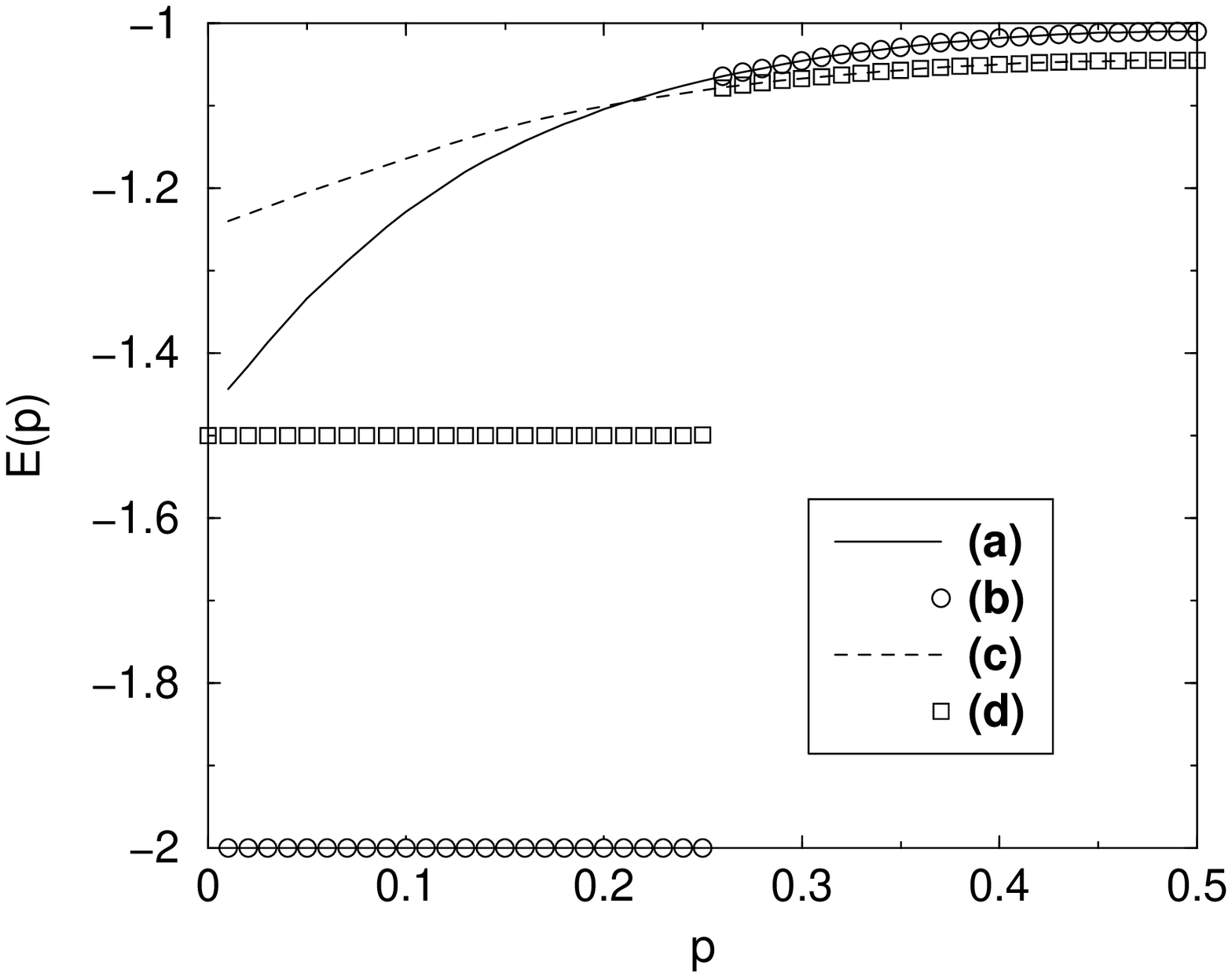}
\caption{}
\label{fig5}
\end{figure}

\begin{figure}
\narrowtext
\epsfxsize=0.8\hsize
\epsfbox{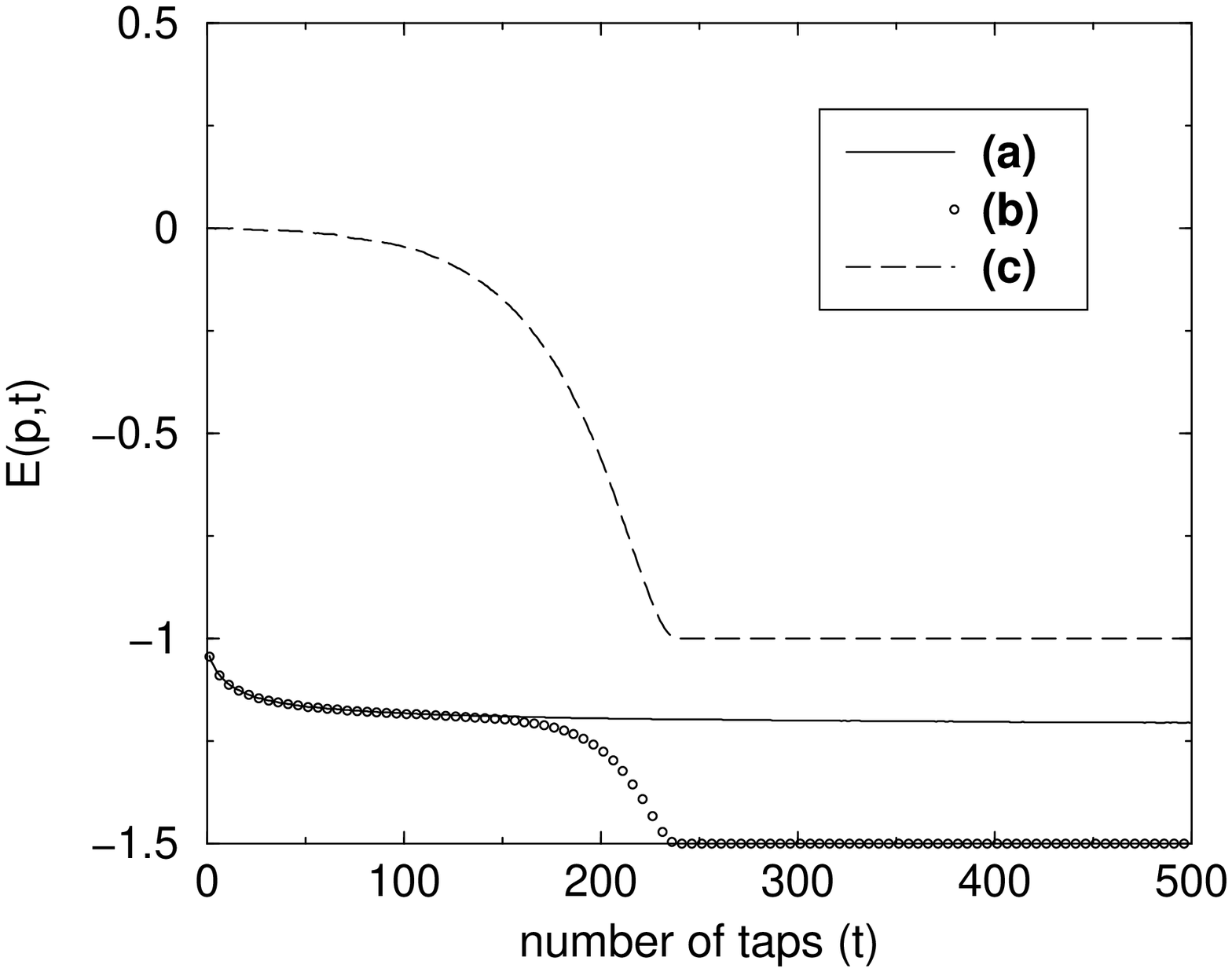}
\caption{}
\label{fig6}
\end{figure}

\begin{figure}
\narrowtext
\epsfxsize=0.8\hsize
\epsfbox{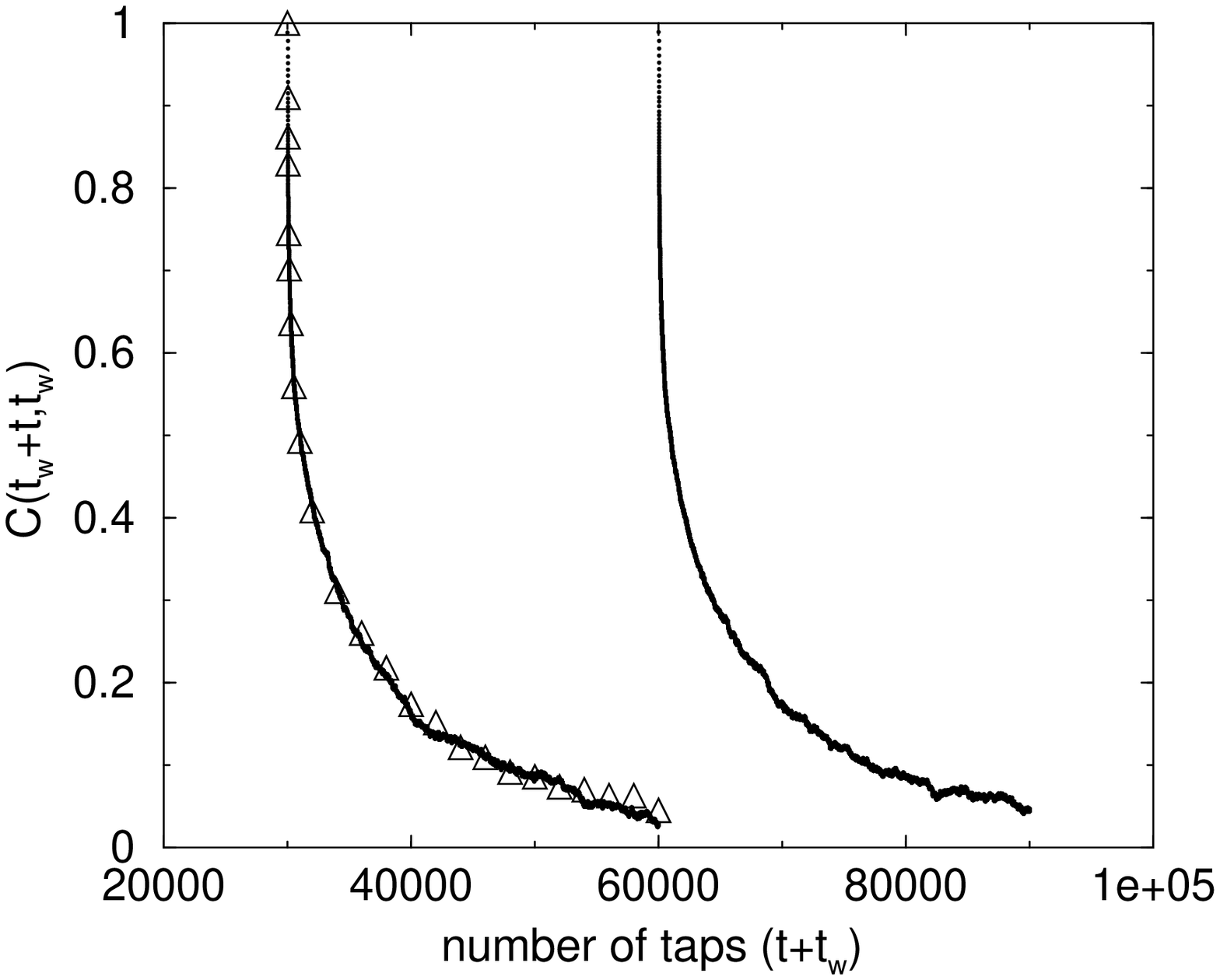}
\caption{}
\label{fig7}
\end{figure}

Fig. 3. Histogram of the energy per spin obtained from Fig.(\ref{fig4}).

Fig. 4. Single run for a ferromagnet of local connectivity c=3, for
        $N=10000$ at  $p=0.249 \approx p_c$. One sees that because 
        the size is not too large, the energy switches between two values,
        one not far from that of the ground state $E_{GS}$ and the other 
        not far from $E^*$.

Fig. 5. Numerical simulations of tapping experiments for $c=3$ ((c): 
        spin glass, (d): ferromagnet) and
	$c=4$ ((a): spin glass, (b): ferromagnet). Here, we have computed 
        the asymptotic energy for only one
	sample of very big size $N=10^6$. The results are quite the same as
	for $N=10^5$ and do not change if we average over several samples,
	which indicates that at these sizes, we are very near the
	thermodynamic limit and we can study only single runs to compute the
	energy.

Fig. 6. Comparison of the energy versus time (number of taps) for the $\pm
        J$ spin glass (a) and the ferromagnet (b) for $N=10^6$ spins at 
        $p=0.05$. We have displayed the magnetization for the ferromagnet 
        (c), whose absolute value increases with time, whereas that of 
        the spin glass remains zero.

Fig. 7. Correlation function for the $\pm J$ spin-glass with local
        connectivity $c=3$ versus number of taps $t$ for two values of the
        waiting time $t_w$: $t_w=30000$ and $t_w=60000$. The tapping value is
        $p=.02$, the system contains $N=1000$ spins and we have averaged over
        $1000$ samples. With  triangles is shown  the right part of the curve,
        which corresponds to $t_w=60000$, shifted to the left by $30000$, it 
        superimposes perfectly over the curve for $t_w=30000$ demonstrating
        the time translation invariance of the correlation function. 
\begin{multicols}{2}
\begin{tabular}{|c|c|c|c|c|c|}
\hline
$c$ & $p_c$ & $E(p_c^+)$ & $E^*$ & $E_f$ & $E_f^{MF}$ \\
\hline
3 & $0.25\pm .005$ & $-1.076\pm .005$ & $-1.0714$ & $-1.045$ & $-1.023$ \\
\hline
4 & $0.255\pm .005$ & $-1.07\pm.005$ & $-1.07\pm .01$ & $-1.01$ & $-1.005$ \\
\hline
5 & $0.45\pm .01$ & $-1.4\pm .005$ & $-1.4\pm .01$ & $-1.396$ & $-1.368$ \\
\hline
\end{tabular}\label{tab1}

\pagestyle{empty}

\end{multicols}
\end{document}